\documentclass[11pt]{article}
\usepackage{moriond,epsfig}

\def\aap{{\em A\&A}}

\def\apj{{\em ApJ}}

\def\prd{{\em Phys.~Rev.~D}}

\def\be{\begin{equation}}
\def\ee{\end{equation}}
\def\bea{\begin{eqnarray}}
\def\eea{\end{eqnarray}}

\def\VEV#1{{\left\langle #1 \right\rangle}}

\def\Mll{M_{\ell \ell'}}
\def\Dl{D_\ell}
\def\Cl{C_\ell}

\begin{document}

{
  \noindent \begin{minipage}{\textwidth}
    \begin{tabular}{p{16cm}}
      \\
      \hline\\
    \end{tabular}
    \begin{center}
    {\bf  To appear in the proceedings of the XXXIXth Rencontres de Moriond}\\
    {\bf  ``Exploring the Universe", La Thuile, Italy, March 28-April
      4, 2004}\\
    \end{center}
    \begin{tabular}{p{16cm}}
      \\
      \hline\\
    \end{tabular}

  \end{minipage}
}

\vspace*{2cm}
\title{A NEW ESTIMATION OF THE ARCHEOPS ANGULAR POWER SPECTRUM}

\author{ M. Tristram}

\address{Laboratoire de Physique Subatomique et de Cosmologie,\\
  53 Avenue des Martyrs, 38026 Grenoble Cedex, France}

\maketitle \abstracts{ We present a refined angular power spectrum of
  the Cosmic Microwave Background (CMB) anisotropies using the {\sc
    Archeops} last flight data. The estimation of the $\Cl$ described
  here is performed using Xspect~\cite{xspect}, a method which uses
  the cross-power spectra of the maps of 6 different detectors. It
  covers multipole range from $\ell=10$ to $\ell=700$ in 25 bins and
  confirms a strong evidence of a plateau followed by the presence of
  two Doppler peaks.  {\sc Archeops} was conceived as a precursor of
  the Planck HFI instrument by using the same optical design and the
  same technology for the detectors and their cooling. Since last
  publication, specific methods have been developed, extra bolometers
  were used and the sky coverage, of about 20\%, is almost twice
  larger. We also present a comparison with first-year WMAP data both
  using individually spectra and by computing the cross-power spectrum
  of the two experiments.}

\section{Introduction}
{\sc Archeops}~\footnote{see {\tt http://www.archeops.org}}, a
pre-WMAP balloon--borne experiment, is a CMB bolometer-based
instrument using {\sc Planck} technology that fills a niche where
previous experiments were unable to provide strong constraints due to
insufficient sky coverage. Namely, {\sc Archeops} seeks to join the
gap in multipole $\ell$ between the large angular scales as measured
by {\sc Cobe}/{\sc Dmr} and degree-scale experiments, typically for
$\ell$ between 10 and 200. These results have been published in
2002~\cite{archeops_cl}.

{\sc Archeops} consists of a 1.5~m aperture diameter telescope and an
array of 21~photometers maintained at $\sim 100$~mK which operates at
4~frequency bands centered at 143, 217, 353 and 545~GHz.  The data
were taken during the Arctic night of February~7,~2002 after the
instrument was launched by CNES from Esrange base (Sweden). The entire
dataset covers $\sim 30$\% of the sky. The refined analysis presented
here was obtained with a subset of the data using the 6 most sensitive
photometers in the CMB bands (143 and 217~GHz) and 19.9\% of the sky.

\section{Data processing\label{processing}}


A detailed description of the instrument inflight performance is given
in~\cite{archeops_instrument} and the full data processing is
described in~\cite{archeops_pipe}. Here we only focus on the new
refined analysis we developed with respect to~\cite{archeops_cl}.


The CMB dipole is the prime calibrator of the instrument. The absolute
calibration error against the dipole measured by
COBE/DMR~\cite{fixsen} and confirmed by WMAP~\cite{wmap} is
estimated to be less than 4\% (resp.~8\%) in temperature at 143~GHz
(resp.~217~GHz).

One of the main assumptions of the power spectrum estimation method is
that the detector noises are decorrelated. So it is capital to
carefully and efficiently remove correlated noise and systematic
effects.  To suppress residual dust and atmospheric signals, the data
are decorrelated using a linear combination of the high frequency
photometers (353 and 545~GHz) and a synthetic dust timeline
extrapolated from SFD observations~\cite{SFD} at~100~$\mu$m.

\section{Beams}
The beam shapes of the bolometers measured on Jupiter are quite
irregular and so the effective beam transfer function for each
bolometer must be carefully estimated. The asymmetry imposes to take
into account the scanning strategy. The beam transfer functions are
computed with simulations using the {\it Asymfast} method detailed in
\cite{asymfast}. This method is based on the decomposition of the beam
into a sum of Gaussians which are easily convolved in the spherical
harmonic space. This allows to deal with effective asymmetric beam
patterns using the scanning strategy of the instrument.

\begin{figure}[!ht]
  \center
  \includegraphics[scale=1.5]{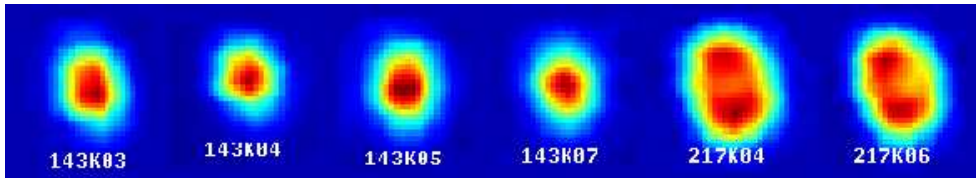}
\caption{The Archeops beam patterns of the six CMB bolometers used in
  this analysis.}
\label{fig_FP}
\end{figure}

Figure~\ref{fig_FP} presents the 6 CMB detectors used in this
analysis. While the 143~GHz detector beams are quite elliptical with
a FWHM of about 11~arcmin, the 217~GHz ones are rather irregular with
a FWHM of about 13~arcmin.

\section{Map-making\label{map} and Galactic mask}
Detector maps of 7~arcmin. resolution (HEALPix~\cite{healpix}
$nside=512$) are produced using an optimal procedure called
MIRAGE~\cite{mirage} for each absolutely calibrated detector with data
bandpassed between 0.1 and 38~Hz. The high-pass filter removes
remaining atmospheric and galactic contamination, the low-pass filter
suppresses non-stationary high frequency noise.

A Galactic mask is deduced from SFD~\cite{SFD} map extrapolated at
353~GHz. The Galactic plane and the Taurus region are efficiently
hiden by asking for an emission $< 0.5$~MJy.sr$^{-1}$. So CMB maps
cover 19.9\% of the sky sampled by $\sim 100,000$ pixels.

\section{Power spectrum estimation, Xspect}
The estimation of the angular power spectrum is performed using an
extension of the so-called `pseudo-$\Cl$'s
estimators~\cite{peebles,spice,master} to the cross power spectrum
called Xspect~\cite{xspect}. The method computes directly the data
{\it `pseudo' angular power spectrum}, $\Dl$, linked to the power
spectrum, $\Cl$ by
\begin{equation}
  \widehat{\Dl^{AB}}
  =
  \sum_{\ell'} \Mll^{AB} p_{\ell'}^2 B_{\ell'}^A B_{\ell'}^B F_{\ell'}^{AB}
  \VEV{C_{\ell'}^{AB}} + N_\ell^{AB}
  \label{pseudo_cross}
\end{equation}
where the mode-mode coupling kernel resulting from the cut sky
$\Mll^{AB}$ is computed analytically for each cross power spectra and
the beam transfer function $B_\ell$ for each bolometer. The $p_\ell$
pixel transfert function takes into account the smoothing effect
induced by the finite size of the pixel. Filtering function
$F_\ell^{AB}$ is deduced from CMB simulations. Assuming no correlation
between the noises of two different maps, the noise term $N_\ell^{AB}$
vanishes and each cross spectrum is an unbiased estimation of the
$C_{\ell}$ (see~\cite{xspect}).

After correction, the cross power spectra are optimally combined into
an accurate $\widetilde{C_\ell}$ estimation using only the `real'
cross spectra ($A \neq B$) which are not noise-biased. The combination
method is based on the maximization of the likelihood,
\begin{equation}
  -2 \ln {\cal L} = \sum_{ij} \left[ (C_\ell^i - \widetilde{\Cl}) |\Xi^{-1}|_\ell^{ij}
    (C_\ell^j - \widetilde{C_\ell}) \right]
\label{likelihood}
\end{equation}
where, $\left| \Xi_\ell \right|^{ij}$ is the analytically computed
cross-correlation matrix of the cross power spectra $C_\ell^i$ and
$C_\ell^j$ (see \cite{xspect} for the description of $\Xi$).

With this method, we can use individual weighting schemes for each
cross power spectra. Two different weighting schemes are combined to
produce the smallest error bars. For the low-$\ell$ part of the
spectrum, each pixel has equal weighting while in the high part the
data are noise weighted (a $1/\sigma^2$ weighting of the data was done
in each pixel, where $\sigma^2$ is the variance of the data in that
pixel).

\begin{figure}[!ht]
  \resizebox{!}{!}{
    \psfig{figure=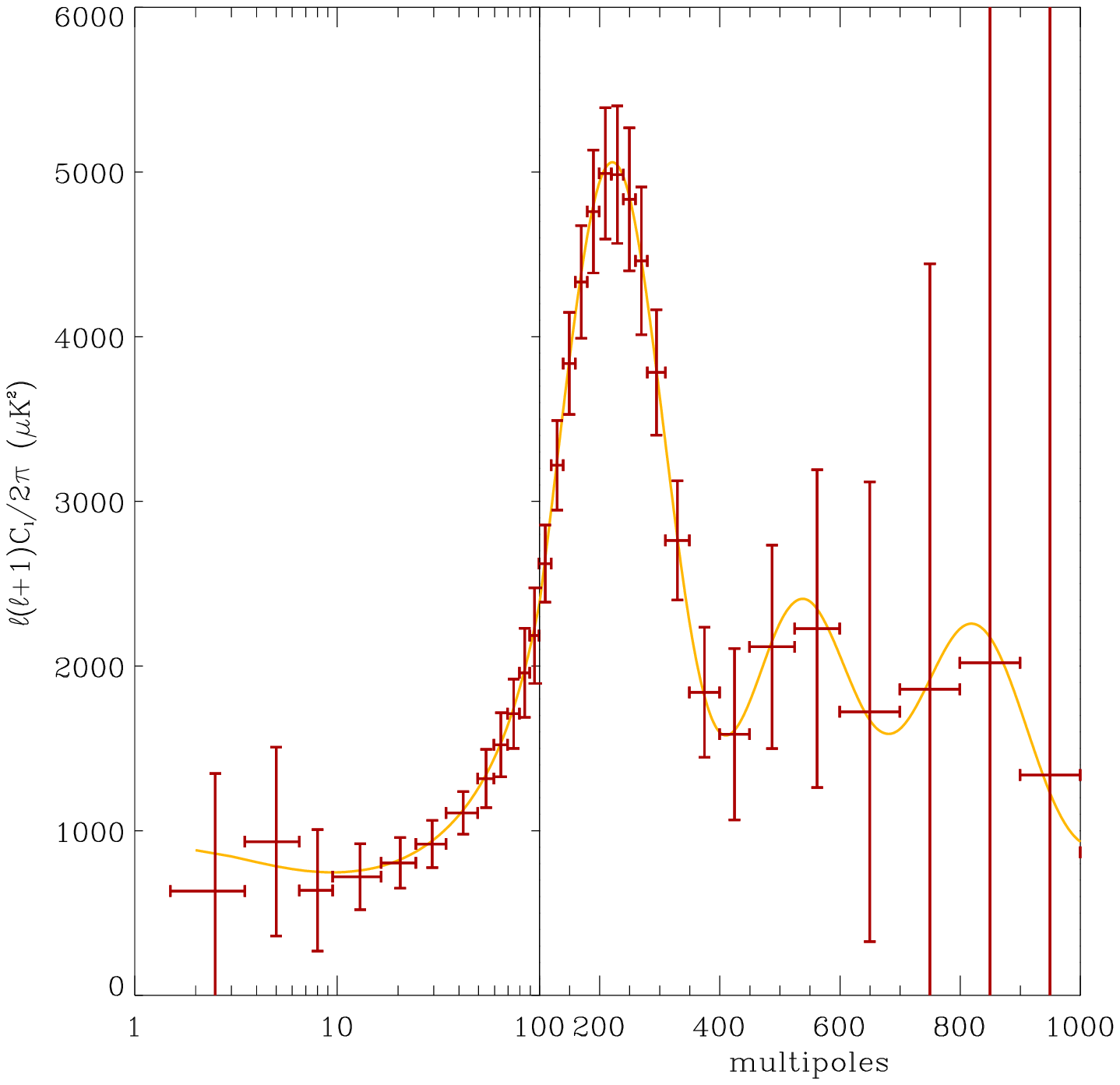,scale=0.4}
    \hspace{0.5cm}
    \psfig{figure=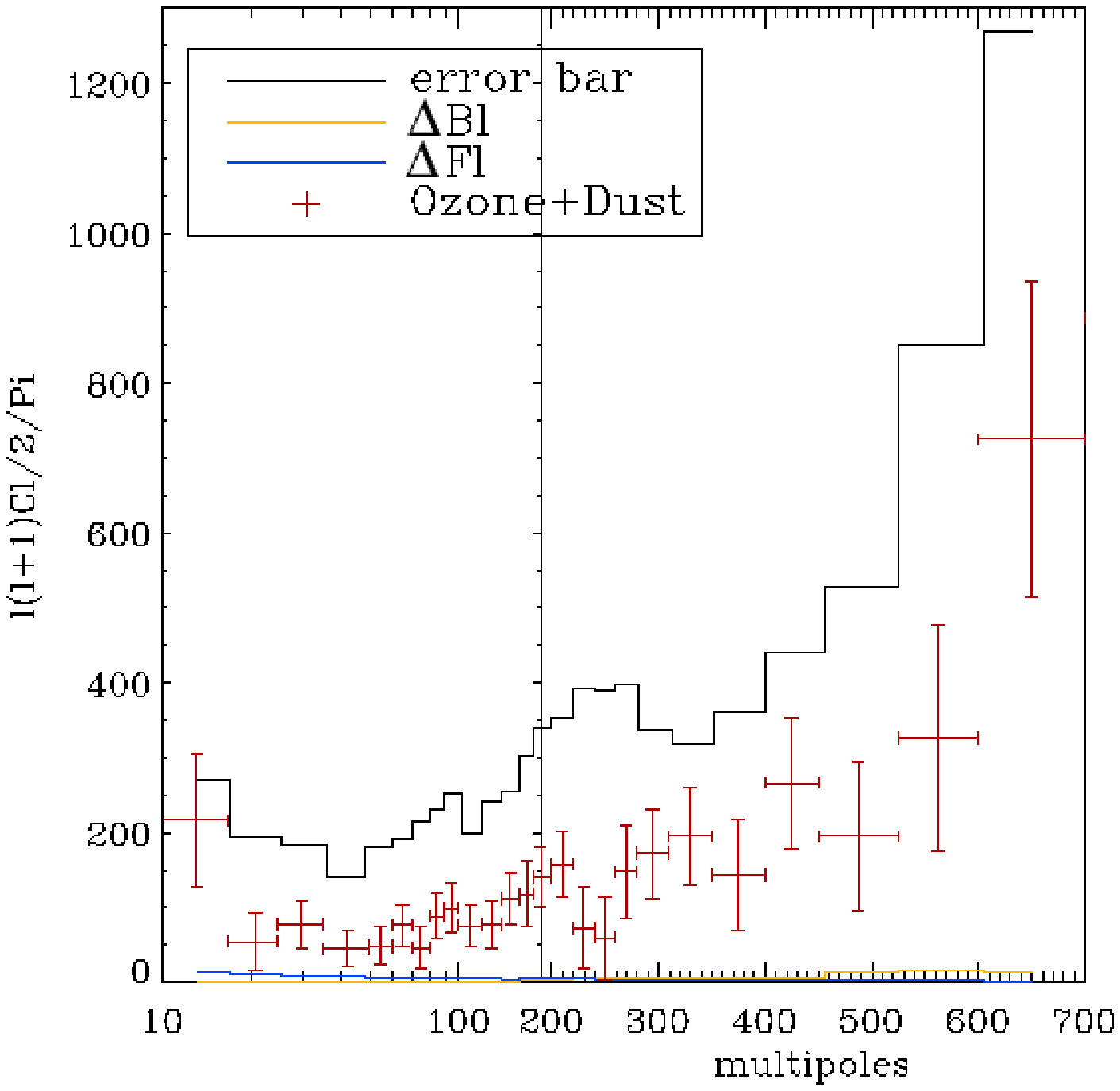,scale=0.4}
  }
  \caption{Left : Power spectrum of 500 realistic simulations of {\sc
      Archeops} data computed using Xspect. Right : Preliminary
    contamination by systematics of {\sc Archeops} data.}
  \label{archcl}
\end{figure}

Figure~\ref{archcl} shows the mean angular power spectrum computed
from 500 simulations of {\sc Archeops} setup with analytically
computed error bars. Simulations are done from realizations of the
CMB sky convolved by the beams and realistic noise is added in
timelines which are filtered with the same bandpass as for the data. We
observe that the optimal combination of cross-power spectra is an
unbiased estimate of the input CMB model for the angular power
spectrum.

Results on the {\sc Archeops} last flight data have been shown on the
conference and will be soon published by the {\sc Archeops}
collaboration. Compared to the last publication~\cite{archeops_cl},
the spectrum presents smaller error bars on a larger multipole range
(from $\ell=10$ to $\ell=700$) using a larger sky coverage and extra
bolometers.

\section{Systematics and Foregrounds contamination}
As a balloon-borne experiment, {\sc Archeops} is exposed to the
atmospheric emission. Moreover the Galactic emission at 143 and
particularly at 217~GHz is low but not negligeable. Even if a carefull
decorrelation of ozone and dust emission has been performed (see
Sect.~\ref{processing}), uncertainties of their residuals in the CMB
data are the major source of systematic errors.

To estimate the level of contamination coming from these foregrounds
we cross-correlate the detectors used for the $C_\ell$ estimation with
the 353~GHz {\sc Archeops} channel. We thus check if there is still
some Galactic and atmospheric contaminations in the CMB maps.
Figure~\ref{archcl} summarizes the different sources of systematic
errors. They remain bellow the sample variance at low multipoles and
the instrumental noise at higher $\ell$.

Tests of consistency have been computed in order to check the
robustness of the results. They consist in computing the power
spectrum using Xspect with different map resolutions, varying
frequency cuts in bandpass filtering or increasing the Galactic mask
coverage. They show a very stable power spectrum.

\section{Comparison with WMAP}
The Xspect method has been applied to the
WMAP~\footnote{http://lambda.gsfc.nasa.gov} data restricted to the {\sc
  Archeops} sky coverage. This was done in order to be able to compare
the power spectrum of the two data sets individually coming from the
same sky area. Taking into account a 8\% calibration uncertainties of
{\sc Archeops} in temperature, we found that the two power spectra are
compatible with $\chi^2$ by dof~=~19.3/24. In parallel, we also
cross-correlate the {\sc Archeops} data with the WMAP ones. {\sc
  Archeops} rescaled by 8\% is correlated using Xspect to WMAP maps.
The cross-power spectrum is found to be compatible with the WMAP one
with $\chi^2$ by dof~=~25.6/24. A more detailed analysis is under
study.

\end{document}